\documentstyle[preprint,aps]{revtex}
\begin{document}

\newcommand{\beq} {\begin{equation}}
\newcommand{\eeq} {\end{equation}}
\newcommand{\pder}[2]{{\partial #1\over\partial #2}}
\newcommand{\chH}{{\hat{\cal H}}}
\newcommand{\chO}{{\hat{\cal O}}}
\newcommand{\chS}{{\hat{\cal S}}}
\newcommand{\hx}{\hat{x}}
\newcommand{\hp}{\hat{p}}
\newcommand{\hr}{\hat{r}}
\newcommand{\hmu}{\hat{\mu}}
\newcommand{\hT}{\hat{T}}
\newcommand{\half}{{1\over 2}}
\newcommand{\fourth}{{1\over 4}}
\newcommand{\eighth}{{1\over 8}}
\newcommand{\arctanh}{\mathrm{\ arctanh}}
\newcommand{\csch}{\mathrm{\ csch}}
\newcommand{\sech}{\mathrm{\ sech}}
\newcommand{\emath}{\mathrm{e}}

\draft
\preprint{\vbox{\hbox{KUNS-1410}\hbox{HE(TH) 96/11}}}
\title{Operator Transformations Between Exactly
  Solvable Potentials and Their Lie Group Generators}
\author{Andrew J. Bordner\thanks{e-mail address: 
bordner@gauge.scphys.kyoto-u.ac.jp} \\
Department of Physics, Kyoto University, Kyoto 606-01, Japan}
\date{\today}
\maketitle
\begin{abstract}
One may obtain, using operator transformations, algebraic relations
between the Fourier transforms of the causal propagators of different
exactly solvable potentials.  
These relations are derived for the shape invariant potentials.
Also, potentials related by real transformation functions are shown to 
have the same spectrum generating algebra with Hermitian generators related by
this operator transformation.
\end{abstract}
\section{Introduction}
The study of exactly solvable potentials, for which the
quantum mechanical eigenfunctions may be expressed in terms of
hypergeometric functions, has a long and varied history.  One approach 
is an algebraic solution of the problem.  Early work by Infeld and
Hull classified factorizations of the Schroedinger operator for solvable
potentials which then allow one to generate other solutions to the
problem. \cite{Infeld_Hull}  A related technique, supersymmetric 
quantum mechanics, discovered 
as a limiting case ($d=1$) of 
supersymmetric field theory, was introduced by Witten and later
developed by other authors.  \cite{SUSY_QM}  In
particular, 
Gendenshtein gave a
criteria, shape invariance, which when satified insures that the
complete spectrum of the supersymmetric Hamiltonian may be found.
\cite{Gendenshtein}  Finally, spectrum generating algebras, whose use 
dates back to Pauli's work on the hydrogen atom, have been studied
more recently as a method to find the spectrum and eigenstates of
solvable potentials.  \cite{Pauli,SGA}

Another method to find the energy eigenvalues and wavefunctions of a
solvable potential is to use an operator transformation, essentially a 
change of independent and dependent variables, to relate it to a
Schroedinger equation for a potential whose solutions are known. 
Duru and Kleinert described such a method for transforming the resolvant
operator, whose matrix element is the propagator. \cite{Duru_Kleinert} 
They used this
technique to transform the time-sliced form of a path integral into a
known path integral, such as that for the harmonic oscillator, by
transforming both the space and time variables in the path integral
expression.  We will discuss these transformations, outside the
context of path integrals, in the next section.

We will show that the operator transformations not only allow one to find
algebraic relations between the Fourier transform of the propagators
for two different quantum systems but also, in the case of a real
transformation function, provide a mapping between the group
generators for the spectrum generating algebra.  Thus quantum systems which may
be mapped to one another by real Duru-Kleinert transformations have
the same formulation in terms of the enveloping algebra of the same
Lie group.

In the first section we describe the operator transformations with
special attention to how the measure for the normalization of states 
transforms.  We next illustrate
the method with a derivation of the relation between the propagators
for the trigonometric Poschl-Teller and Rosen-Morse potentials and 
give the relations for the propagators for some other exactly solvable 
potentials.  Finally we examine the corresponding
transformation of the Lie group generators.

\section{Operator Transformations and Causal Green's Functions}
We will consider transformations of the Fourier transform of the
causal propagator for a quantum mechanical system.  Hereafter
operators 
will be denoted by a caret.  The propagator is
given by 
\beq
K(x_{0},x_{f},t) \equiv \theta(t) \: \langle x_{f} | 
e^{-{\imath \over \hbar} \chH t} | x_{0} \rangle
\eeq
and its Fourier transform is defined by
\begin{eqnarray} 
\label{E_prop}
G(x_{0},x_{f},E) &\equiv& \imath \int_{-\infty}^{\infty}dt\: 
e^{{\imath \over \hbar}E t} \: K(x_{0},x_{f},t) \\ \label{propagator}
&=&  \imath \int_{0}^{\infty}dt \: \langle x_{f} | 
e^{-{\imath \over \hbar} (\chH-E) t} | x_{0} \rangle \nonumber\\ 
&=& \langle x_{f} | {\hbar \over {\chH - E - \imath \epsilon}} 
| x_{0} \rangle \nonumber
\end{eqnarray}
where the infinitesimal imaginary constant in the last line gives the causal
propagator.  

Duru and Kleinert realized that Eqn. \ref{E_prop} is invariant
under two types of operator transformations.  
One type is simply a point canonical transformation, which for a
one-dimensional system is
\begin{eqnarray}
\hx \rightarrow f(\hx) \\ \label{point_CT}
\hp \rightarrow {1 \over f'(\hx)}\hp \nonumber
\end{eqnarray}
with $\hx$, $\hp$ the canonical position and momentum respectively.
This point canonical transformation may be implemented by a similarity 
transformation on the operators, which is also called a quantum canonical
transformation, since if it is applied to all operators it preserves 
the canonical commutation
relations.\cite{Anderson}  Under such a similarity transformation 
\begin{eqnarray}
\chH - E &\rightarrow& \chO (\chH - E) \chO^{-1} \\ \label{similarity}
\langle x |  &\rightarrow& \langle x | \chO^{-1}. 
\end{eqnarray}
The operator $\chO$ which implements the transformation is composed of
the canonical position and momentum operators.  We will assume that
$\chO$ is invertible although, with proper care, operators with a
nonzero kernel may also be considered. \cite{Anderson}  
Clearly, this type of transformation leaves invariant any matrix
element of an operator.  

Another type of transformation which leaves Eqn. \ref{propagator}
invariant is what Duru and Kleinert denoted as an f-transformation.  
We will distinguish between two types of f-transformations, since the
normalization measure transforms differently in each case.  The first
type of f-transformation is a similarity transformation with $\chO =
f(\hx)$, where $f(q)$ is some function of q.  The other type of
transformation, which we will call conjugation, is 
\begin{eqnarray}
\chH - E &\rightarrow& f(\hx) (\chH - E) f(\hx) \\ \label{conjugation}
\langle x |  &\rightarrow& \langle x | f(\hx).
\end{eqnarray}
Eqn. \ref{propagator} is invariant under this transformation, however
a general matrix element of an operator is not invariant.

We next examine the change in the measure factor for these
transformations.  First consider a similarity transformation, 
Eqn. \ref{similarity}.  The original wavefunction $\psi(r)$ and the
transformed wavefunction $\psi'(r)$ are defined as 
\begin{eqnarray}
\psi(r) = \langle r | \psi \rangle \\
\psi'(r) = \langle r | \chO | \psi \rangle 
\end{eqnarray}
with $\langle r |$ an eigenstate of the position operator with
eigenvalue $r$.  We then may find the transformation of the (in
general operator valued) measure factor $\hmu$.
\begin{eqnarray}
\langle \psi | \psi \rangle_{\hmu} &=& \int dr \langle \psi | \hmu | r \rangle 
\langle r | \psi \rangle \\
&=& \int dr \langle \psi | \hmu \chO^{-1}| r \rangle 
\langle r | \chO | \psi \rangle \nonumber\\   
&=& \int dr \langle \psi | \chO^{\dagger} (\chO^{-1})^{\dagger} \hmu 
\chO^{-1}| r \rangle \langle r | \chO | \psi \rangle \nonumber\\ 
&=& \int dr \langle \psi | \chO^{\dagger} \hmu' | r \rangle \langle r | \chO |
\psi \rangle \nonumber\\ 
&=& \langle \psi' | \psi' \rangle _{\hmu'}.\nonumber
\end{eqnarray}
Therefore the measure factor for the transformed wavefunctions is 
$\hmu' = (\chO^{-1})^{\dagger} \hmu \chO^{-1}$. 

We next assume that the measure factor contains only the position operator,
{\it i.e.}, $\hmu = g(\hx)$.  Without ambiguity we may then use the 
notation $g(r)$ for 
the measure factor.  For a point canonical transformation,
Eqn. \ref{point_CT}, the measure transforms as a differential 
\beq
g(r)  \rightarrow g(f(r)){df(r) \over dr}. \label{PCT_measure}
\eeq
For the similarity transformation with $\chO = f(\hx)$ the measure
factor transforms multiplicatively as
\beq
g(r) \rightarrow f^{-2}(r)g(r).
\eeq
Finally the measure factor remains unchanged for the conjugation
transformation of Eqn. \ref{conjugation}.

\section{Example: Rosen-Morse to Poschl-Teller Potential}
The transformation from a Hamiltonian with potential $V_{0}(r)$ 
\beq
\chH = {\hp^{2}\over {2\mu}} + V_{0}(\hx)
\eeq
to another with potential $V_{f}(r)$ is specified by a single function $f(r)$.
We will illustrate the general sequence of transformations along with
the specific example with $V_{0}(r)$ the Rosen-Morse I potential and
$V_{f}(r)$ the hyperbolic Poschl-Teller potential.    

First a point canonical transformed is performed as in
Eqn. \ref{point_CT}.  For the example, the function is $f(r) = {1\over
  a}\arctanh \cos 2ar$, giving the operator transformation
\begin{eqnarray}
\hx &\rightarrow& \chO_{0}\hx\chO_{0}^{-1}= {1\over a}\arctanh \cos 2a\hx \\
\hp &\rightarrow& \chO_{0}\hp\chO_{0}^{-1}= -\half \left(\sin
  2a\hx\right) \hp, \nonumber
\end{eqnarray}
which transforms the original operator, $\chS_{0} \equiv \chH_{0} - E$ 
\beq
\chS_{0} = {1\over 2\mu}\hp^{2} + A\csch^{2} a\hx + B\coth ax \csch ax 
- E
\eeq
into 
\beq
\chS_{1} \equiv \chO_{0}\chS_{0}\chO_{0}^{-1} = {1\over 8\mu} 
\left( \sin^{2} 2a\hx \hp^{2}- 2a\imath \hbar \sin 2a\hx 
\cos 2ax \hp \right) + A \cos 2a\hx - B \sin^{2} 2a\hx -E.
\eeq
According to Eqn. \ref{PCT_measure} the measure transforms as 
\beq
dx \rightarrow {-2\over \sin 2a\hx} dx.
\eeq
The propagator becomes 
\begin{eqnarray}
G_{\mathrm{R-M}}(x_{f},x_{0},E) &=& \imath \int dT \langle x_{f} |\: 
e^{-{\imath\over \hbar}\chS_{0}T}\: | x_{0}\rangle \\
&=& \imath \int dT \langle x_{f} |\: \chO_{0}^{-1} e^{-{\imath\over 
\hbar}\chS_{0}T} \left(\chO^{-1}\right)^{\dagger}\: | x_{0}\rangle \\ \nonumber
&=& \imath \int dT \langle {1\over 2a}\arccos(\tanh ax_{f}) |\: 
e^{-{\imath\over \hbar}\chS_{1}T} \:| {1\over 2a} \arccos(\tanh ax_{0}) 
\rangle. \nonumber
\end{eqnarray}
Next one performs the similarity transformation with $\chO_{1} =
\left({df(r)\over dr}\right)^{\half}=\sin^{-\half} 2a\hx$ to get
\begin{eqnarray}
\chS_{2} \equiv \chO_{1}\chS_{1}\chO_{1}^{-1} &=& {1\over 8\mu}\sin^{2}
2a\hx \hp - {\imath\hbar a\over 2\mu}\sin 2a\hx \cos 2a\hx \hp 
+ {2\hbar^{2}a^{2}\over {8\mu}} \sin^{2} 2a\hx \\ 
&+& A \cos 2a\hx - B \sin^{2}
2a\hx - {\hbar^{2}a^{2}\over {8\mu}} - E. \nonumber
\end{eqnarray}
The measure transforms as 
\beq
{-2\over \sin 2a\hx} dx \rightarrow -2 dx
\eeq
and the propagator is then
\begin{eqnarray}
G_{\mathrm{R-M}}&(&x_{f},x_{0},E) = \imath \int dT \langle {1\over 2a}\arccos(\tanh
ax_{f}) |\: \chO^{-1}
e^{-{\imath\over\hbar}\chS_{2}T}\left(\chO^{-1}\right)^{\dagger}\: |
{1\over 2a}\arccos(\tanh ax_{0})\rangle \\
&=&\imath (\sech^{\half} ax_{f})(\sech^{\half} ax_{0}) \int dT 
\langle {1\over 2a}\arccos(\tanh
ax_{f}) |\: e^{-{\imath\over\hbar}\chS_{2}T}\:
|{1\over 2a}\arccos(\tanh ax_{0})\rangle. \nonumber 
\end{eqnarray} 

Next a conjugation transformation follows, Eqn. \ref{conjugation}, with the
function $C{df(r)\over dr}$.  The constant $C$ is chosen to give the 
correct kinetic energy factor in the Hamiltonian.
\begin{eqnarray}
\chS_{3} \equiv {2\over \sin 2a\hx}\chS_{2}{2\over \sin 2a\hx} &=& 
{1\over 2\mu}\hp^{2} + \left(A - E - {{\hbar^{2}a^{2}}\over 
{8\mu}}\right) \csc^{2} 2a\hx \\
&+& \left(-A - E - {{\hbar^{2}a^{2}}\over
{8\mu}}\right) \sec^{2} 2a\hx -\half\hbar^{2}a^{2} - 4B. \nonumber
\end{eqnarray}
The transformed propagator is 
\begin{eqnarray}
G_{\mathrm{R-M}}&(&x_{f},x_{0},E) = \imath (\sech^{\half} ax_{f})
(\sech^{\half} ax_{0}) \\ \label{RM_propagator}
&\times& \int dT \langle {1\over 2a}\arccos(\tanh
ax_{f}) |\:\left({2\over \sin 2a\hx}\right)
e^{-{\imath\over\hbar}\chS_{3}T} \left({2\over \sin 2a\hx}\right)
\:|{1\over 2a}\arccos(\tanh ax_{0})\rangle \nonumber\\  
&=& 4 \imath(\cosh^{\half} ax_{f}(\cosh^{\half} ax_{0}) \nonumber\\ 
&\times&\int dT \langle {1\over 2a}\arccos(\tanh
ax_{f}) |\: 
e^{-{\imath\over\hbar}\chS_{3}T}\:| {1\over 2a}\arccos(\tanh
ax_{f})\rangle. \nonumber
\end{eqnarray} 
Finally the the Hilbert space is rescaled so that the measure becomes
the usual one, $\mu = dx$, 
\beq
^{\mathrm{norm}}\langle x | \equiv \sqrt{2} \langle x |.
\eeq
This introduces a factor of $\half$ in the propagator,
Eqn. \ref{RM_propagator}.  The final result is
then obtained
from Eqn. \ref{RM_propagator} by matching parameters in the operator
$\chS_{3}$ with those for the Poschl-Teller potential.
The algebraic relations between the Fourier transform of the
propagator for several solvable potentials are shown in the table
along with 
the function $f(r)$ used for the operator transformations.
\footnote{The transformation functions given in the table are also listed in 
Ref. \cite{point_CT_map}, however we correct them for the Rosen-Morse
II and Eckart potentials.} 
Although all of the potentials for which we give explicit results in
the table are shape invariant, the operator transformations are valid
for a general potential.  It is interesting to note that although not
all one dimensional
solvable potentials, classified by Natanzon, are shape invariant, they 
are related to a shape invariant potential by an operator
transformation.  \cite{Natanzon,solvable_SUSY}

\section{Operator Transformations for Lie Group Generators}
The operator transformations from $\chS_{0} \equiv \chH_{0} - E_{0}$ to 
$\chS_{f} \equiv \chH_{f} - E_{f}$ may be summarized by 
\beq
\label{S_trans}
\chS_{f} = C\:(f')^{3/2}\:\chO_{0}\:\chS_{0}\:\chO_{0}^{-1}\:(f')^{\half}.
\eeq
$\chO_{0}$ is the operator implementing the point canonical
transformation, Eqn. \ref{point_CT}, with function $f(q)$ and C is a
constant.  Since the
eigenvalue equation, $\chS_{f}=0$, is homogeneous one may multiply
Eqn. \ref{S_trans} by $C^{-1}(f')^{-2}$ on the left to obtain the following 
equation, valid for 
an interval in which $f'\neq 0$ and finite,
\beq
\label{L_trans}
(f')^{-\half}\:\chO_{0}\:\chS_{0}\:\chO_{0}^{-1}\:(f')^{\half}=0.
\eeq
The operator transformation between the eigenvalue equation for the
Hamiltonian $\chH_{0}$ and $\chH_{f}$ now preserves the commutators 
of operators on the
two Hilbert spaces, {\it e.g.}, it is a Lie algebra isomorphism. 
The new generators $\hT_{f}^{i}$ are related to the Lie
algebra generators for the original potential, $\hT_{0}^{i}$ as 
\beq
\label{gen_trans}
\hT_{f}^{i}=(f')^{-\half}\:\chO_{0}\:\hT_{0}^{i}\:\chO_{0}^{-1}\:(f')^{\half}
\eeq 
Therefore, in the cases where the
eigenvalue equation for $\chH_{0}$ may be written as an element of the 
enveloping
algebra of a particular Lie algebra, the transformed eigenvalue
equation, Eqn. \ref{L_trans}, has the same formulation in terms of Lie 
group generators, however in a different representation.  The 
eigenvalue equation for the
potentials listed in the table
then have the same Lie algebraic form as either the radial harmonic
oscillator, the trigonometric Poschl-Teller, or the hyperbolic
Poschl-Teller potential.  $SU(1,1)$ generators for the radial
harmonic oscillator Schroedinger operator and those related to it by
Eqn. \ref{gen_trans} are well known and given in Ref. \cite{so21_algebra}.    

As an example, we consider the Lie algebraic form for the
trigonometric Poschl-Teller potential and then find the transformed
generators for the Rosen-Morse I potential.
The Poschl-Teller potential is known to have an algebraic formulation
in terms of the Lie group $SU(2) \otimes SU(2)$.  
One may find the generators for $SO(4)=SU(2) \otimes SU(2)$ by
considering the generators of rotations in $\Re^{4}$ 
\begin{eqnarray}
J_{1} &=& {\imath\over 2}\left(-x_{1}\partial_{4} + x_{2}\partial_{3} -
  x_{3}\partial_{2} + x_{4}\partial_{1}\right), \\
J_{2} &=& {\imath\over 2}\left(-x_{1}\partial_{3} - x_{2}\partial_{4} +
  x_{3}\partial_{1} + x_{4}\partial_{2}\right), \nonumber \\ 
J_{3} &=& {\imath\over 2}\left(-x_{1}\partial_{2} + x_{2}\partial_{1} +
  x_{3}\partial_{4} - x_{4}\partial_{3}\right), \nonumber\\ 
K_{1} &=& {\imath\over 2}\left(-x_{1}\partial_{2} + x_{2}\partial_{1} -
  x_{3}\partial_{4} + x_{4}\partial_{3}\right), \nonumber\\
K_{2} &=& {\imath\over 2}\left(x_{1}\partial_{3} - x_{2}\partial_{4} -
  x_{3}\partial_{1} + x_{4}\partial_{2}\right), \nonumber\\
K_{3} &=& {\imath\over 2}\left(x_{1}\partial_{4} + x_{2}\partial_{3} -
  x_{3}\partial_{2} - x_{4}\partial_{1}\right). \nonumber
\end{eqnarray}

Changing to Euler angle coordinates for the double cover of $S^{3}$
\begin{eqnarray}
x_{1} &=& \cos\left({\theta\over 2}\right)
\cos\left({{\phi+\psi}\over 2}\right), \\
x_{2} &=& \cos\left({\theta\over 2}\right)
\sin\left({{\phi+\psi}\over 2}\right), \nonumber\\
x_{3} &=& \sin\left({\theta\over 2}\right)
\cos\left({{\phi-\psi}\over 2}\right), \nonumber\\ 
x_{4} &=& \sin\left({\theta\over 2}\right)
\sin\left({{\phi-\psi}\over 2}\right), \nonumber
\end{eqnarray}
and scaling $\theta \rightarrow 2a\theta$ we obtain the generators
\begin{eqnarray}
\label{PT_generators}
J_{1} &=& \imath\left({1\over{2a}}\sin{\psi}\partial_{\theta} 
- \csc2a\theta \cos\psi \partial_{\phi} 
+ \cot2a\theta \cos\psi\partial_{\psi}\right), \\ 
J_{2} &=&
\imath\left(-{1\over{2a}}\cos{\psi}\partial_{\theta} 
-\csc 2a\theta \sin\psi \partial_{\phi} 
+ \cot 2a\theta \sin \psi \partial_{\psi}\right), \nonumber\\ 
J_{3} &=& -\imath\partial_{\psi}, \nonumber\\ 
K_{1} &=& \imath\left({1\over{2a}}\sin{\phi}\partial_{\theta} 
+ \cot2a\theta \cos\phi\partial_{\phi}  
- \csc2a\theta \cos\phi \partial_{\psi}\right), \nonumber\\ 
K_{2} &=&
\imath\left(-{1\over{2a}}\cos{\phi}\partial_{\theta} 
+ \cot 2a\theta \sin \phi \partial_{\phi} 
-\csc 2a\theta \sin\phi \partial_{\psi}\right), \nonumber\\
K_{3} &=& -\imath\partial_{\phi}. \nonumber
\end{eqnarray}
These obey the commutation relations
\begin{eqnarray}
\left[J_{l},J_{m}\right] &=& \imath\:\epsilon_{lmn}\:J_{n}, \\
\left[K_{l},K_{m}\right] &=& \imath\:\epsilon_{lmn}\:K_{n}, \nonumber\\
\left[J_{l},K_{m}\right] &=& 0, \nonumber
\end{eqnarray}
and $J_{i}$ is obtained from $K_{i}$ by interchanging $\phi
\leftrightarrow \psi$.  These operators are similar to those found in
Ref. \cite{Barut}, which were deduced from the corresponding
Infeld-Hull factorization.  The Casimir operator $J^{2}$ is 
\begin{eqnarray}
4a^{2}J^{2} &=& -\partial^{2}_{\theta} + a^{2}\left(-\partial^{2}_{\phi} -
\partial^{2}_{\psi} + 2\partial_{\phi}\partial_{\psi} - \fourth\right)
\csc^{2} a\theta \\
&+& a^{2} \left(-\partial^{2}_{\phi} -
\partial^{2}_{\psi} - 2\partial_{\phi}\partial_{\psi} -
\fourth\right)\sec^{2}a\theta - a^{2}. \nonumber
\end{eqnarray}
The other Casimir operator $K^{2}$ is identical.  
One may express the eigenfunction equation for a unitary
representation of the group $SU(2)$ as 
\begin{eqnarray}
\label{su2_rep}
J^{2}|klm\rangle &=& k(k+1)|klm\rangle, \ k=0,\half,1,{3\over 2},\ldots \\
J_{3}|klm\rangle &=& n|klm\rangle, \ n = -k,\ldots,0,\ldots,k. \nonumber
\end{eqnarray}
If one chooses the eigenfunction $|klm\rangle=u_{mn}^{k}(\theta)
\emath^{\imath(l\phi+m\psi)}$ then Eqn. \ref{su2_rep} becomes
\begin{eqnarray}
{{2a^{2}}\over\mu}&J&^{2}u_{lm}^{k}(\theta) \\
&=& \left[-{1\over{2\mu}}
\partial^{2}_{\theta} 
+ {a^{2}\over{2\mu}}\left((l-m)^{2}-\fourth\right)\csc^{2}a\theta 
+ {a^2\over{2\mu}}\left((l+m)^{2}-\fourth\right)\sec^{2}a\theta 
- {a^{2}\over{2\mu}}\right]
u_{lm}^{k}(a\theta) \nonumber \\
 &=& {{2a^{2}}\over \mu}k(k+1)u_{lm}^{k}(\theta). \nonumber
\end{eqnarray}
This is the Schroedinger equation for the Poschl-Teller potential,
which if we define the coefficients in the potential 
$A\equiv \hbar^{2}\gamma(\gamma-1)$ and 
$B\equiv \hbar^{2}\delta(\delta-1)$, gives $\gamma = l-m+\half$,
$\delta=l+m+\half$ and $E_{k} = {{2a^{2}\hbar^{2}}\over
    \mu}(k+\half)^{2}$.  Since $l = k - j,\ j = 0,1,\ldots,2k$ the
  energy eigenvalues are
\beq
E_{k} = {{a^{2}\hbar^{2}}\over{2\mu}}(\gamma+\delta+2j)^{2}
\eeq
with $j\geq \half(1-\gamma-\delta)$.
The same procedure for the $K_{i}$ operators gives the same energy
eigenvalues.  

If one transforms the $SU(2)$ generators $J_{i}$, in
Eqn. \ref{PT_generators}, into the corresponding ones for the
Rosen-Morse I potential, using Eqn. \ref{gen_trans}, one obtains
\begin{eqnarray}
J^{\mathrm{RM}}_{1} &=& \imath\left({-1\over a}\cosh a\theta 
  \sin \psi \partial_{\theta} - \cosh a\theta \cos \psi
  \partial_{\phi} + \sinh a\theta \cos \psi \partial_{\psi}\right), \\
J^{\mathrm{RM}}_{2} &=& \imath\left({1\over a}\cosh a\theta 
  \cos \psi \partial_{\theta} -\cosh a\theta \sin \psi \partial_{\phi} 
+\sinh a\theta \sin \psi \partial_{\psi}\right), \nonumber \\
J^{\mathrm{RM}}_{3} &=& -\imath\partial_{\psi}. \nonumber
\end{eqnarray}
The Casimir operator acting on the state $|klm\rangle \equiv 
u_{lm}^{k}(\theta)\emath^{\imath(l\phi-m\psi)}$ gives
\begin{eqnarray}
J^{2}u_{lm}^{k}(\theta) &=& \left[{{-\cosh^{2}
      a\theta}\over a}
\partial_{\theta}^{2} + (l^{2}+m^{2})\cosh^{2} a\theta 
+2lm \sinh a\theta \cosh a\theta \right]u_{lm}^{k}(\theta) \\
&=& k(k+1)u_{lm}^{k}(\theta) \nonumber 
\end{eqnarray}
and $J^{\mathrm{RM}}_{3}|klm\rangle = -n|klm\rangle$.
Multiplying by $-a^{2}\hbar^{2}\sech^{2} a\theta/2\mu$ leads to 
the Schroedinger
  equation for the Rosen-Morse potential 
\beq
\hbar^{2}\left[-{1\over 2\mu}\partial_{\theta}^{2} +
  {{a^{2}lm}\over{\mu}}\tanh a\theta 
- {{a^{2}k(k+1)}\over{2\mu}}\sech^{2} a\theta \right]u_{lm}^{k}(\theta) 
= -{{a^{2}\hbar^{2}(l^{2}+m^{2})}\over{2\mu}}u_{lm}^{k}(\theta)
\eeq
with parameters $A=a^{2}lm/\mu$ and $B=a^{2}k(k+1)/2\mu$ and energy eigenvalue 
$E = -a^{2}\hbar^{2}(l^{2}+m^{2})/2\mu$.  Since the energy eigenvalues 
are non-positive only the bound states energies may be found.  Again
for a unitary representation of
$SU(2)$ we have $-m = -k + j,\ j=0,1,\ldots,2k$.  Substituting this
in the equation for the energy eigenvalue and expressing the result in 
terms of the potential coefficients 
\begin{eqnarray}
E_{j} &=& -\hbar^{2}\left[{\mu A^{2}\over{2a^{2}}}\left({1\over n^{2}}\right) 
+{a^{2}\over{2\mu}}n^{2}\right] \\
n &=& -\half + \half\sqrt{1 + {{8\mu B}\over a^{2}}}-j,\ j = 0,1,\ldots,
\left(-1+\sqrt{1+{{8\mu B}\over a^{2}}}\right). \nonumber
\end{eqnarray}
Furthermore we may assume that $A\geq0$, since under the change of
variables $\theta \rightarrow -\theta$, $A \rightarrow -A$.  Similar 
to the Poschl-Teller case, the other $SU(2)$ operators $K_{i}$ may be
found from $J_{i}$ by exchanging $\phi \leftrightarrow \psi$ 
and furthermore the
Casimirs are equal, 
$K^{2}=J^{2}$.  Therefore, with $K_{3}|klm\rangle = l |klm\rangle$ 
the range of the eigenvalue is $l = -k,-k+1,\ldots,k-1,k$ and one
finds the following bound on the coefficients in the potential in order for the
existence of a bound state 
\beq
\left({\mu A\over{a^{2}}}\right)^{\half} = lm \leq k^{2} = \left(-\half +
  \half\sqrt{1+{{8\mu B}\over a^{2}}}\right)^{2}.
\eeq
\section{Conclusion}
We have shown that if a particular type of operator transformation,
which is not
necessarily unitary, exists between two Schroedinger operators there
is a procedure for finding an algebraic relation between the
respective propagators and that the two eigenvalue problems have the
same formulation in terms of Lie group generators.  Also a knowledge of
the Fourier transform
of the propagator for the new potential allows one, in principle, to
find the energy eigenvalues and wavefunctions for both the bound and
scattering states.  
One interesting generalization of this procedure would be to find such 
operator transformations between multiparticle exactly solvable systems,
such as those of the Calogero-Sutherland type.  
\section*{Acknowledgements}
This work was supported by the Japanese Society for the Promotion of Science.

\end{document}